\documentclass[a4paper]{article}

\usepackage{INTERSPEECH2022}
\usepackage{INTERSPEECH2022}
\usepackage{amsmath,graphicx}
\usepackage{comment}
\usepackage{cite}
\usepackage{caption, color}
\usepackage{adjustbox}

\usepackage[labelformat=simple]{subcaption}

\title{An Empirical Analysis on the Vulnerabilities of End-to-End Speech Segregation Models}
\name{Rahil Parikh$^1$, Gaspar Rochette$^2$, Carol Espy-Wilson$^1$, Shihab Shamma$^1$}

\address{
  $^1$Institute for Systems Research\\ University of Maryland College Park, USA\\
  $^2$ENS Paris, PSL Université, France}
\email{rahil@umd.edu}

\begin{document}

\maketitle
\begin{abstract}
End-to-end learning models have demonstrated a remarkable capability in performing speech segregation. Despite their wide-scope of real-world applications, little is known about the mechanisms they employ to group and consequently segregate individual speakers. Knowing that harmonicity is a critical cue for these networks to group sources \cite{parikh2022harmonicity}, in this work, we perform a thorough investigation on ConvTasnet \cite{luo2019conv} and DPT-Net \cite{chen2020dual} to analyze how they perform a harmonic analysis of the input mixture. We perform ablation studies where we apply low-pass, high-pass, and band-stop filters of varying pass-bands to empirically analyze the harmonics most critical for segregation. We also investigate how these networks decide which output channel to assign to an estimated source by introducing discontinuities in synthetic mixtures.
We find that end-to-end networks are highly unstable, and perform poorly when confronted with deformations which are imperceptible to humans. Replacing the encoder in these networks with a spectrogram leads to lower overall performance, but much higher stability. This work helps us to understand what information these network rely on for speech segregation, and exposes two sources of generalization-errors. It also pinpoints the encoder as the part of the network responsible for these errors, allowing for a redesign with expert knowledge or transfer learning.


\end{abstract}
\noindent\textbf{Index Terms}: end-to-end speech segregation, Conv-Tasnet, harmonic continuity, temporal consistency, generalization
\vspace*{-1mm}
\section{Introduction}
\label{sec:intro}
Speech communication often occurs in complex acoustic environments with concurrent sounds. To realize speech processing applications such as Automatic Speech Recognition, Speaker Verification and Identification, etc. in such environments, the system must be capable of segregating speech of individual speakers from a mixture of speakers.
Advances in deep learning have facilitated drastic improvements in single-channel speech segregation \cite{hershey2016deep, zhang2016deep, luo2018tasnet, luo2019conv, luo2020dual, chen2020dual, kolbaek2017multitalker, huang2015joint, isik2016single, chen2017deep, erdogan2015phase, shi2019deep}. 
Some of these models \cite{luo2018tasnet, luo2019conv} outperform the Ideal Ratio Mask (IRM) \cite{narayanan2013ideal} when trained and evaluated on the WSJ-2-Mix dataset \cite{hershey2016deep}, allowing them to be used in a wide variety of downstream applications in the \textit{wild}. These advances have been adopted to other source segregation tasks such as Universal Source Segregation \cite{kavalerov2019universal} and Music Segregation \cite{defossez2019music}, directly affecting an even wider array of downstream audio tasks, such as Sound Event Detection \cite{Wisdom_InPrep2020, Turpault2019_DCASE}. 

\begin{figure}
  \centering
  \includegraphics[width=\columnwidth]{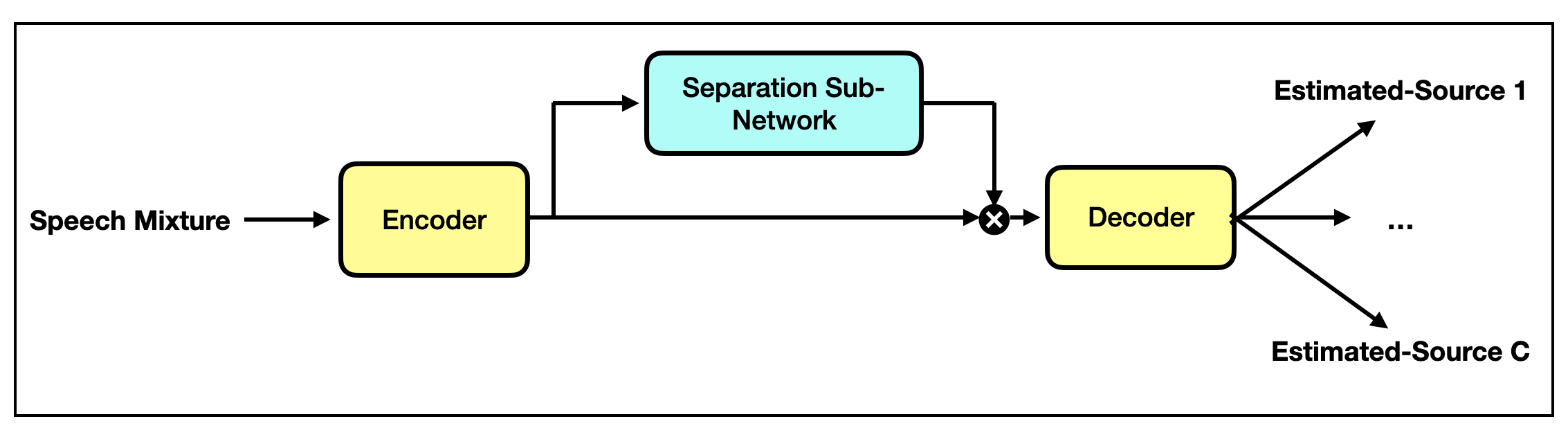}
  \caption{A typical E2E speech segregation network design}
  \label{fig:end-to-end DNNS}
\end{figure}

End-to-end (E2E) speech-segregation systems are believed to out-perform spectrogram based models \cite{luo2018tasnet}. A typical design is shown in Fig \ref{fig:end-to-end DNNS}, where the model is trained to segregate a mixture of $C$ speakers into $C$ individual streams. The separation sub-net generates $C$ masks on the encoded mixture waveform. The element-wise product of the encoded mixture and masks are decoded to produce each estimated speaker. The encoder, separation sub-network and decoder are trained end-to-end to optimize the Scale Invariant Source-to-Noise Ratio (SI-SNR) or Signal-to-Distortion Ratio (SDR) \cite{le2019sdr, mcfee2015librosa, vincent2006performance}  using Permutation Invariant Training (PIT) \cite{kolbaek2017multitalker}.

\begin{figure}[t]
  \centering
  \includegraphics[width=0.7\columnwidth]{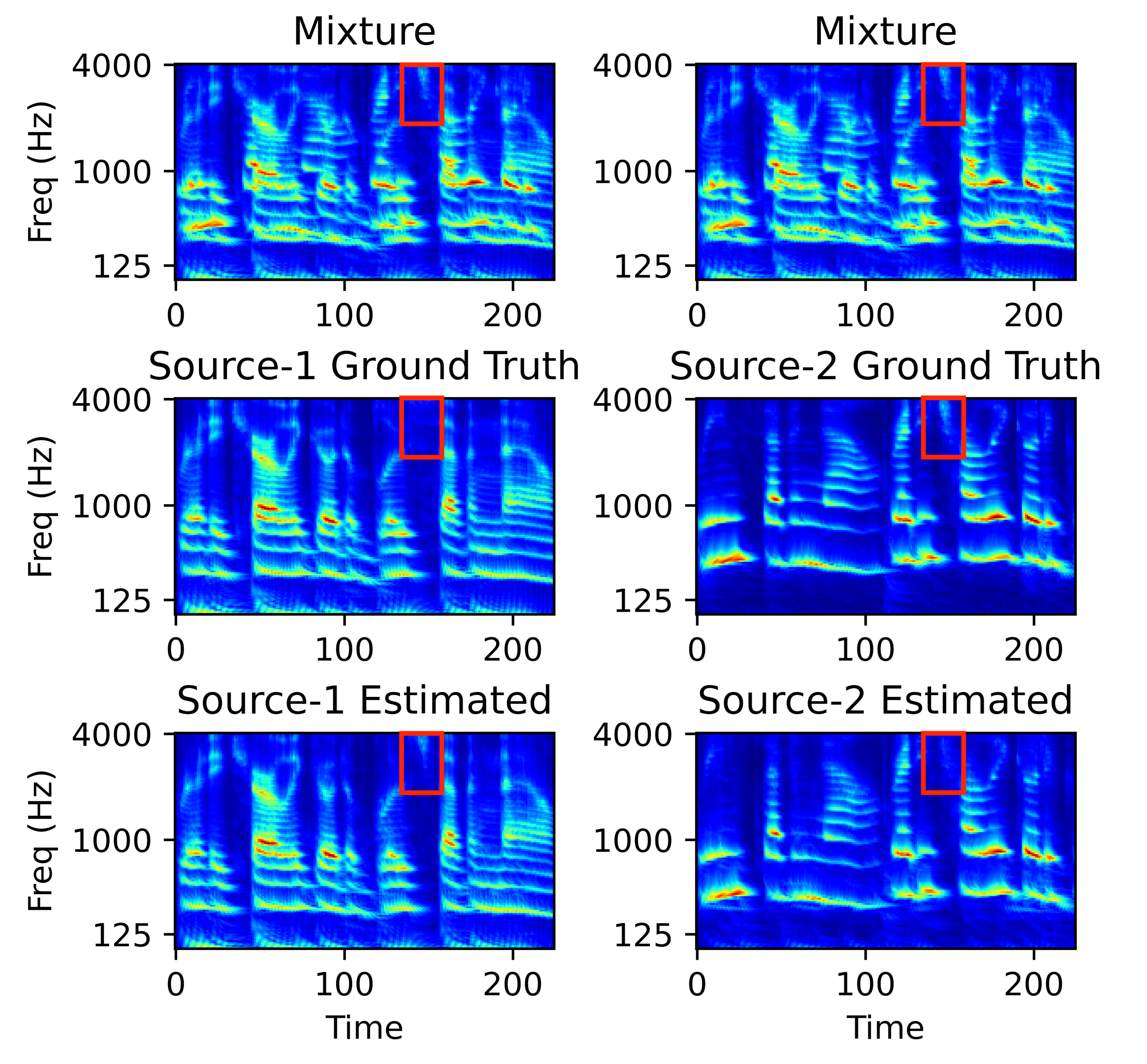}
  \caption{ConvTasnet fails to segregate fricatives in natural speech. The red boxes highlight where ‘/s/’ is grouped with Source-1 instead of Source-2.}
  \label{fig:fricative}
\end{figure}

Due to the black-box nature of deep-learning models, little is known about the principles employed by these models to segregate sources. The sensitivity of E2E speech-segregation models to mixtures of inharmonic speech is demonstrated in \cite{parikh2022harmonicity}, indicating that these networks are heavily reliant on harmonicity to group and segregate sources. This is corroborated by our observation that despite the abundance of fricatives in the training data, ConvTasnet often fails to segregate fricatives in natural speech as demonstrated in Fig \ref{fig:fricative}, although, this error is usually not heard due to the robustness of our speech perception.
Appreciating the importance of harmonicity for segregation, we empirically investigate the factors responsible for these models to produce a harmonic analysis of the mixture and subsequently track these harmonic patterns to eventually perform segregation.

The goal of our paper is to analyze the shortcomings of these
E2E models to better understand the principles responsible for their success. We believe that this would help us - 1) understand the current limitations of these networks, 2) interrelate traditional Computational Auditory Scene Analysis (CASA) algorithms which rely on engineering \cite{ weiss2010speech, cooke2010monaural}, biological \cite{elhilali2008cocktail, krishnan2014segregating}, or pitch tracking \cite{vishnubhotla2009algorithm,stark2010source, wang1999separation} based solutions with E2E networks, and 3) improve these networks by training them to be invariant to these shortcomings.
We observe that-
\begin{itemize} 
\vspace*{-0.5mm}
    \item Speech Segregation is more reliant on cues from the lower harmonics of speech than the higher harmonics.
    \vspace*{-3.5mm}
    \item E2E source segregation models are very sensitive to a discontinuous harmonic space. These networks fail when any intermediate harmonic is absent.
    \vspace*{-1mm}
    \item Short discontinuities of ~20ms in the audio cause a Causal ConvTasnet to perform incorrect assignment near the point of discontinuity
    \vspace*{-1mm}
    \item Spectrogram-based models trained to generate masks directly on the mixture spectrogram are more robust to harmonic deformations and temporal discontinuities. 
\end{itemize}
\vspace*{-4mm}
\section{Experiments}
We perform our analysis on ConvTasnet and DPT-Net trained to segregate 2 speakers. First, we investigate whether the models attend to the higher harmonics which are more prone to overlapping or to the lower harmonics which generally contain more energy.
We then analyze the importance of continuity in the harmonic structure for grouping the channels of each estimated speaker. 
Next, we investigate the importance of temporal continuity to analyze how the models approach the assignment problem of deciding which estimated speaker should be assigned to which channel. 
Lastly, we contrast these results with speech-segregation networks trained directly on spectrograms \cite{chi2005multiresolution}.
\subsection{Datasets and Evaluation Metrics}
We perform our analysis on ConvTasnet and DPT-Net trained on the WSJ-2-Mix data \cite{hershey2016deep}. We report experiments on both, mixtures of speech from the WSJ-2-Mix test-set and mixtures of non-overlapping alternating tones which are more controllable.

We demonstrate the network's behaviour visually by generating mixtures of two alternating, non-overlapping harmonic tones, where each tone has $N$ harmonics and is given by $x_{i}(t)=\sum_{k=1}^{N}a_{k}\sin(2\pi kf_0^{(i)}t), i\in \{1,2\}$. A mixture of these tones containing all the $N$ harmonics can be successfully segregated by the networks. 
It has been shown in previous work that ConvTasnet relies very strongly on the harmonic structure \cite{parikh2022harmonicity}. Input stimuli of such nature allow us to focus on this harmonic structure and investigate how the network analyses it, by removing certain harmonics such as in Sec. \ref{sec:HPF and LPF Speech} and Sec \ref{sec:Continuity_in_harmonics} and precisely inserting discontinuities in Sec \ref{sec:continuity_in_time}.
To perform our analysis on speech, we filter the validation set of the WSJ-2-Mix dataset using low-pass, high-pass or band-stop-filters to analyse the network's performance when missing the lowest, highest or intermediate harmonics.
We evaluate the ConvTasnet model using the SDR. 
\vspace*{-2mm}
\subsection{Results}
\subsubsection{Model Performance on Low-Passed and High-Passed Filtered Speech}
\label{sec:HPF and LPF Speech}
\vspace*{-1mm}
We investigate whether a network trained on natural speech attends more to the lower or higher harmonics for segregation. We low-pass (LP) filter the WSJ-2-mix dataset with cut-off frequencies at 300Hz, 700Hz, and 1200Hz and evaluate a model with this filtered data to analyze if the network can detect the harmonic structure necessary for segregation \cite{parikh2022harmonicity} in the absence of the higher harmonics. Similarly, we also high-pass (HP) filter the dataset, at cut-off frequencies 180Hz, 300Hz, 400Hz and 700Hz and evaluate the model on this filtered data. We baseline these results against networks trained and evaluated on the corresponding filtered dataset. An alternate baseline is to segregate natural speech using the model trained on natural speech, and filtering the estimated sources. In both cases, we observe performances of around 15db and thus only report the former.

\begin{figure}[t]
  \centering
  \includegraphics[width=\columnwidth]{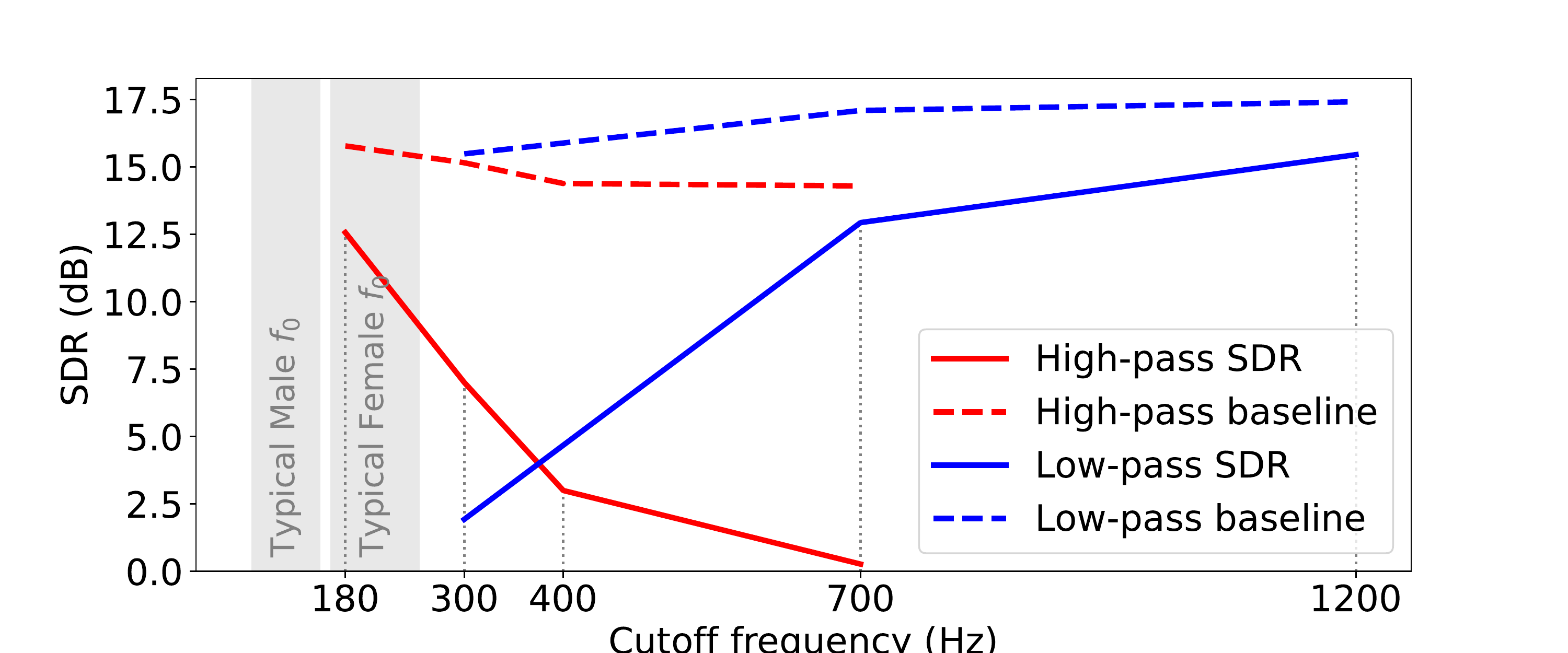}
  \caption{Performance of ConvTasnet on either LP (blue) or HP (red) filtered speech for different cutoff frequencies, compared to the performance of networks trained on filtered data.}
  \label{fig:higher_lower_speech}
\end{figure}

These results are illustrated in Fig \ref{fig:higher_lower_speech}, where it is evident that the network is very sensitive to speech missing its lowest harmonics (e.g. telephonic speech). The network is more robust to the LP filtered data, indicating that the model relies on cues in the lower harmonics more than in the higher harmonics to characterize the incoming mixture.

\begin{figure}[h]
  \centering
  \includegraphics[width=\columnwidth]{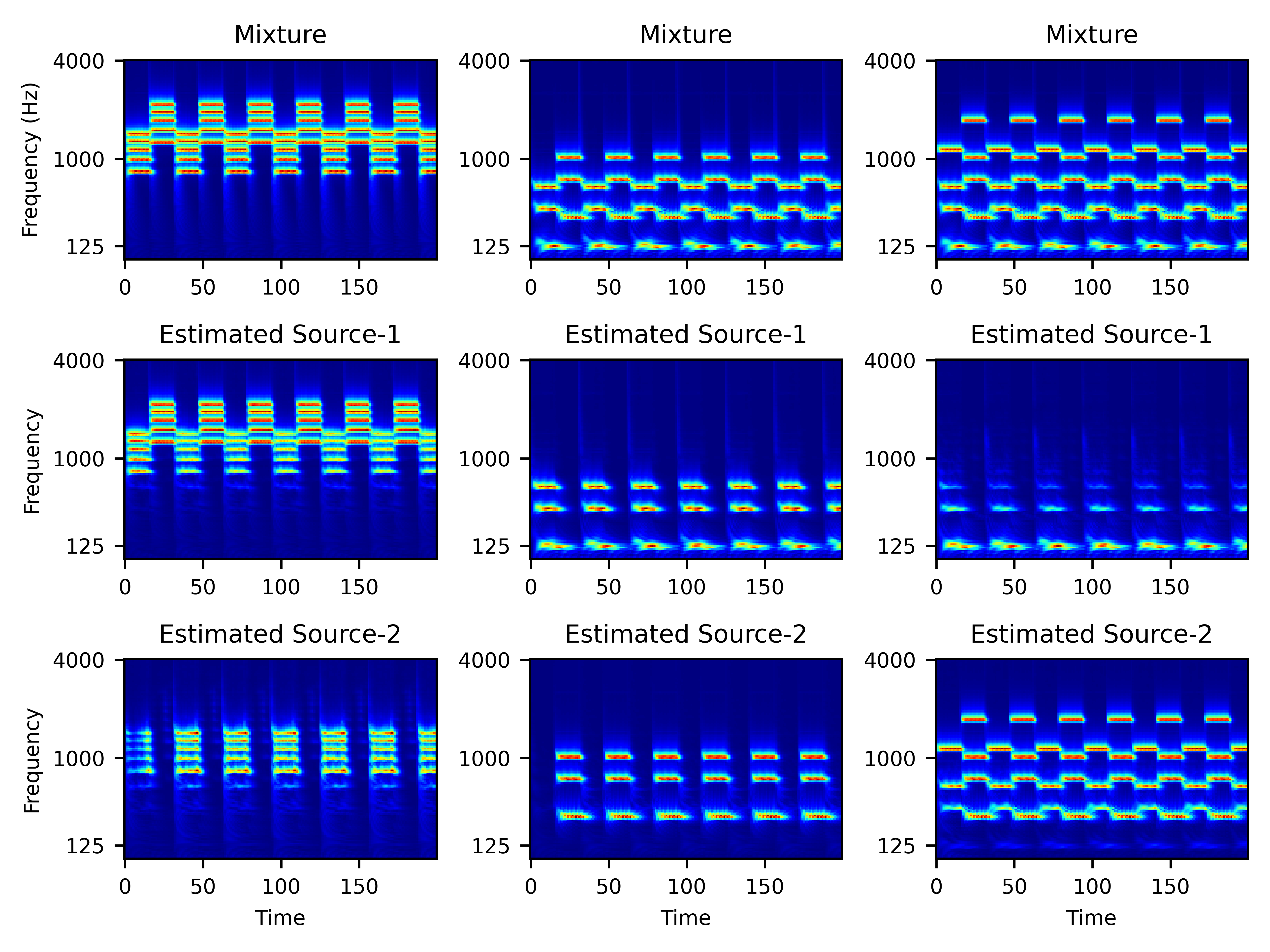}
  \caption{Conv-Tanet fails to segregate mixtures of higher harmonics (left columns) but successfully segregates mixtures of the first three harmonics (centre columns). It  fails to segregate mixtures of the first three and fifth harmonics (right columns)}
  \label{fig:higher_lower_continuity}
  \vspace*{-2mm}
\end{figure}

These results are corroborated by our observations in Fig. \ref{fig:higher_lower_continuity} on mixtures of two alternating tones with F0 as 117Hz and 201Hz. The left column indicates the model's inability to segregate a mixture of tones with the harmonics: $4, \cdots, 8$ and the middle column indicates the model's ability to segregate a mixture with the lowest three harmonics.

\subsubsection{Model Performance on Band-Stopped Filtered Speech}
\label{sec:Continuity_in_harmonics}
\begin{figure}[t]
  \centering
  \includegraphics[width=\columnwidth]{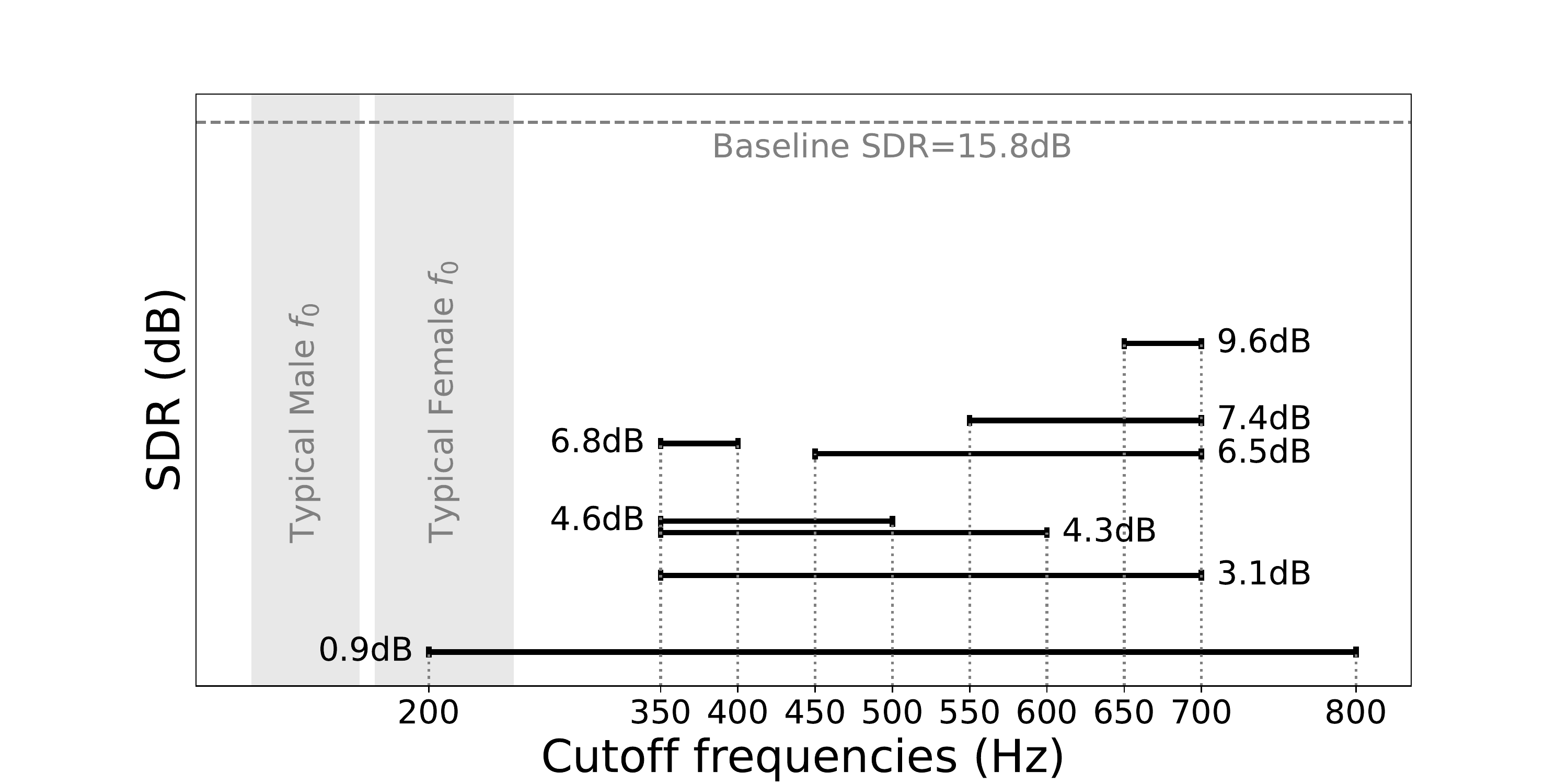}
  \caption{Performance of ConvTasnet on band-stoped data. Each filter is a horizontal line representing the frequencies in the stop-band, and the corresponding SDR.}
  \label{fig:band_stop}
\end{figure}

\vspace*{-1mm}
We investigate the network's sensitivity to discontinuities in the harmonic structure. We generate speech missing a few intermediate harmonics by applying a band-stop filter to the mixture before doing the segregation. We perform this evaluation with a set of 8 band-stop filters, with stop-band frequencies ranging from 200Hz-800Hz, to 350Hz-400Hz. We report the results in Fig. \ref{fig:band_stop} in which each horizontal line represents a band-stop filter by its low and high cutoff frequencies, and reports the corresponding SDR. As in \ref{sec:HPF and LPF Speech}, multiple baselines can be considered which all resulted in performances above 15dB. For clarity, we only report the ConvTasnet native performance of 15.8dB in Fig. \ref{fig:band_stop}.
We repeat this study on the mixture of tones in Fig \ref{fig:higher_lower_continuity}. While the network can segregate a mixture of only the first 3 harmonics (2\textsuperscript{nd} column), it fails when we introduce the 5\textsuperscript{th} harmonic (3\textsuperscript{rd} column). While doing so brings extra information, it is perceived as a missing 4\textsuperscript{th} harmonic which challenges the network's expectations. Unsure of the harmonic structure's position, the network fails and segregates both tones into the same source, indicating its dependence to a continuous harmonic structure.
\vspace*{-2mm}
\subsubsection{Performance on Mixtures with Temporal Discontinuities}
\label{sec:continuity_in_time}
\vspace*{-1mm}
We study the temporal consistency of the network's output by introducing short silences in the data.
\begin{figure}
  \centering
  \includegraphics[width=\columnwidth]{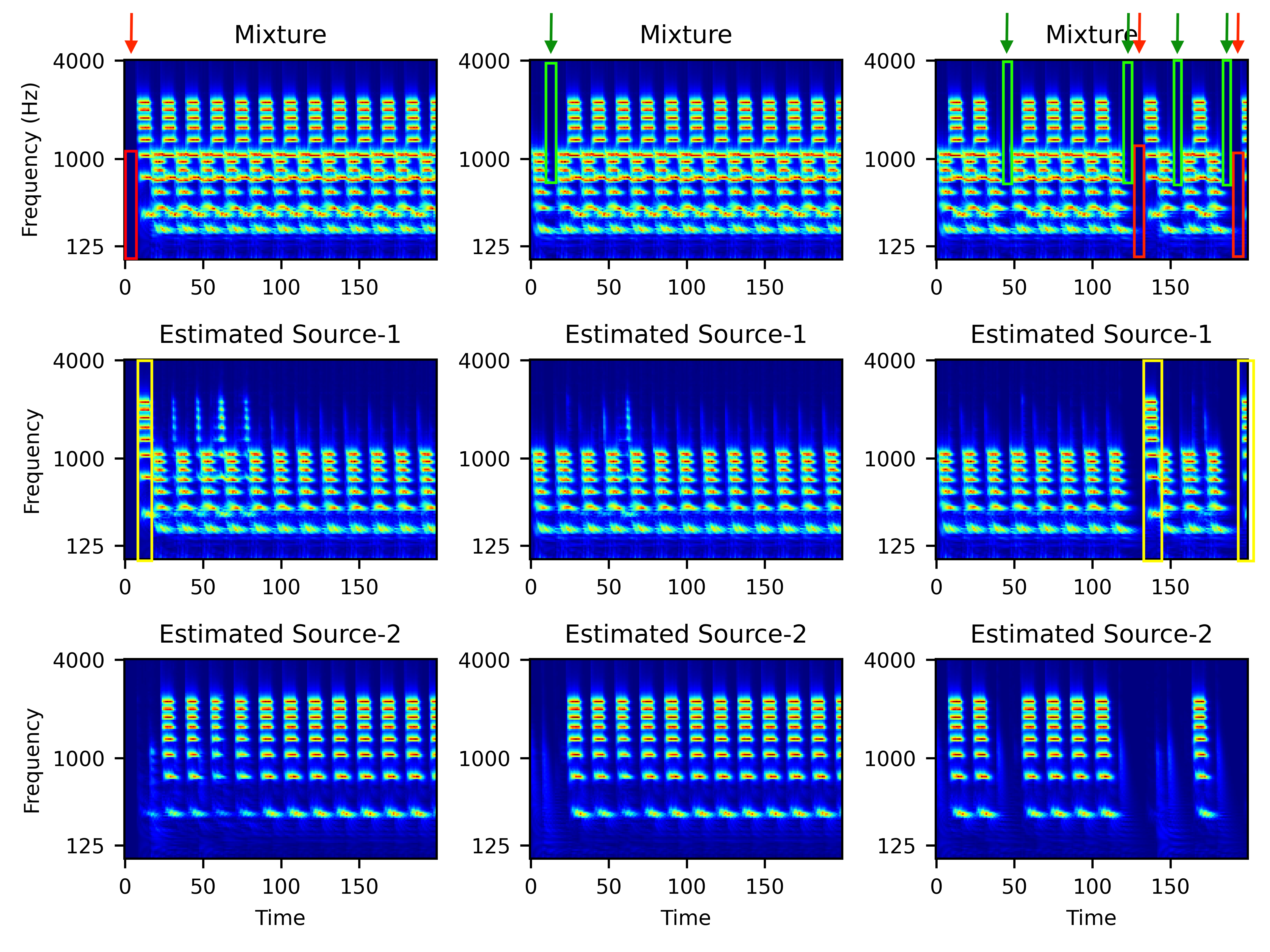}
    \caption{Discontinuities are inserted into a mixture of alternating tones (upper rows). Errors in segregation can be observed when these discontinuities mute the lower frequency tone (yellow boxes in the middle rows).}
  \label{fig:temporal_continuity}
\end{figure}
To visualize the quality of segregation at the point of discontinuity, we use a causal ConvTasnet for these experiments. We create mixtures of two alternating and insert silence ranging from 15ms to 100ms since such silence is usually present in speech, on which the network is trained. Fig \ref{fig:temporal_continuity} shows our results when tone period is 62ms (i.e. 31ms of energy) and the inserted silence (discontinuities) is 31ms in duration. This silence is inserted to the mixture at the points of the arrows in the upper row. The red (respectively green) arrows indicate that the silence has muted the low (respectively high) frequency tone. The corresponding red and green arrows in the mixture spectrograms (upper-rows) indicate the missing tone. 
We observe that the inserted discontinuity leads to an error in segregation, denoted by the yellow boxes in the estimated spectrograms. These errors are localized to the neighborhood of discontinuity, after which the network begins to segregate the sources correctly. More interestingly, these errors only arise when the low frequency tone is muted (red boxes in the mixture spectrogram). Discontinuities that arise from muting the higher frequency tones do not seem to cause an error in segregation. This observation is consistent across tones of periods ranging from 30ms to 125ms and the duration of the discontinuities as transient as 10ms, repeated using 10 random seeds. This behavior is discussed in Sec \ref{sec:Discussion}.
\vspace*{-2mm}
\subsubsection{Comparison of E2E Models with Spectrogram Based Models}
\label{sec:spec_training}
\begin{figure}[t]
  \centering
  \includegraphics[width=0.85\columnwidth]{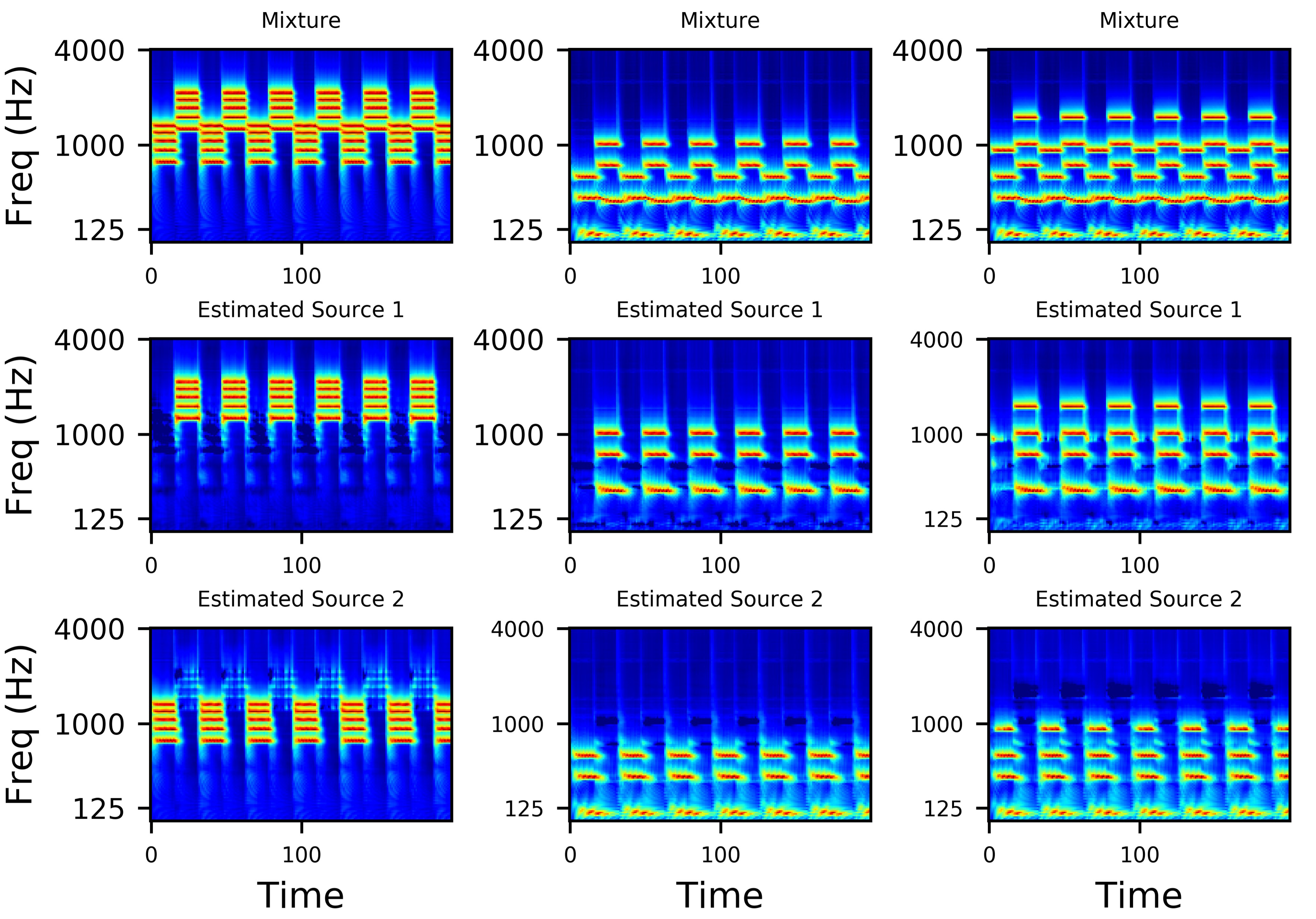}
  \caption{A spectrogram-based network can successfully segregate mixtures of the higher (left column), first three (centre column) and first three and fifth (right column) harmonics}
  \label{fig:spectrogram_training}
\end{figure}

\begin{figure}[h]
  \centering
  \includegraphics[width=\columnwidth]{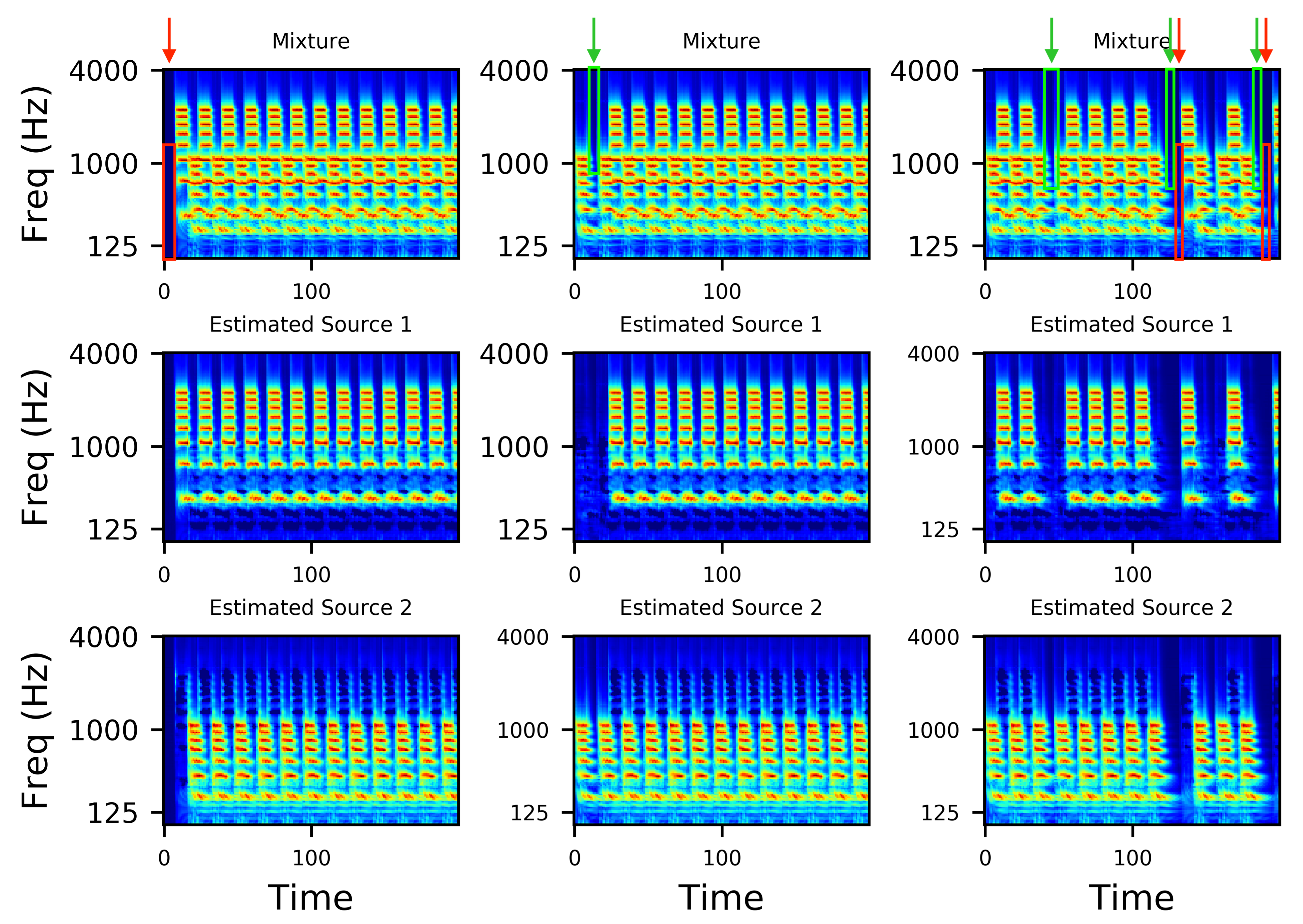}
  \caption{A spectrogram-based model can successfully segregate mixtures with inserted silences which are denoted by red and green arrows in the upper rows.}
  \label{fig:spec_temporal_continuity}
\end{figure}

\vspace*{-2mm}
We analyze the behavior of spectrogram-based speech segregation models on tones with the same harmonic and temporal deformations. 
We train the separation sub-network of ConvTasnet in Fig \ref{fig:end-to-end DNNS} to directly generate masks on the spectrogram. This network is trained to optimize the L2 norm between the estimated and true spectrograms using PIT and is also trained on the WSJ-2-mix dataset. It performs equivalently to a state-of-art spectrogram-based model \cite{kolbaek2017multitalker} on natural speech. 
As illustrated in Fig \ref{fig:spectrogram_training}, when presented with the same stimuli as that in Fig \ref{fig:higher_lower_continuity}, we observe that the spectrogram based model is not sensitive to missing harmonics tests. It can successfully segregate mixtures of just the higher harmonics (left column), lower harmonics (centre column), and mixtures with missing harmonics (right column). 
We also analyze the sensitivity of spectrogram-based models to temporal discontinuities and present the model with the same stimuli discussed in Sec \ref{sec:continuity_in_time}. Fig \ref{fig:spec_temporal_continuity} illustrates that the spectrogram based-model is robust to these deformations and can successfully segregate mixtures with periods of silence.
Our results indicate that spectrogram-based models are significantly more robust to deformations in the evaluation data, in-spite of being trained on the same dataset. 

\vspace*{-2mm}
\subsection{Discussion}
\label{sec:Discussion}
We have demonstrated that ConvTasnet is very sensitive to transformations on the harmonic structure. Removal of the lower harmonics (HP filtering), higher (LP filtering) or intermediate harmonics (band-stop filtering) often occur in the \textit{wild}. These transformations result in an imperceptible deformation to humans but have a catastrophic impact on the network. We repeat our experiments on DPT-Net and observe similar results.
\vspace*{-5mm}
\subsubsection{Sensitivity of Networks to Deformations in Harmonics}
\vspace*{-0.5mm}
Section \ref{sec:HPF and LPF Speech} indicates that E2E segregation networks cue on to the lower harmonics to characterize the harmonic structure. A possible explanation for this is that the lower harmonics have more energy and are often well separated, even in a mixture. Computing the F0 from the higher harmonics is less trivial. For example, energy at 250Hz could either be the F0 of a female speaker, or the second harmonic of a male speaker whose F0 is 125Hz. However, energy at 900Hz could be the 4\textsuperscript{th} or 5\textsuperscript{th} harmonic of a female speaking at respectively 225 or 180Hz, or the 6\textsuperscript{th}, 7\textsuperscript{th}, \dots, or 10\textsuperscript{th} harmonic of a male speaking at respectively 150Hz, 128Hz, \dots, or 90Hz. However, the collection of higher harmonics are perceptually important since humans can segregate speech composed of just the higher harmonics and compute the F0 \cite{shamma2000case}.
Our experiments in Sec \ref{sec:Continuity_in_harmonics} show that even removing a small frequency band from $350-400$Hz or $650-700$Hz considerably degrades the network's performance, from above $15$dB to $6.8$dB and $9.6$dB respectively. This effect is stronger when removing lower frequencies than higher frequencies, and when increasing the range of the stop-band.
\vspace*{-1.5mm}
%

\begin{figure}
  \centering
  \includegraphics[width=0.8\columnwidth]{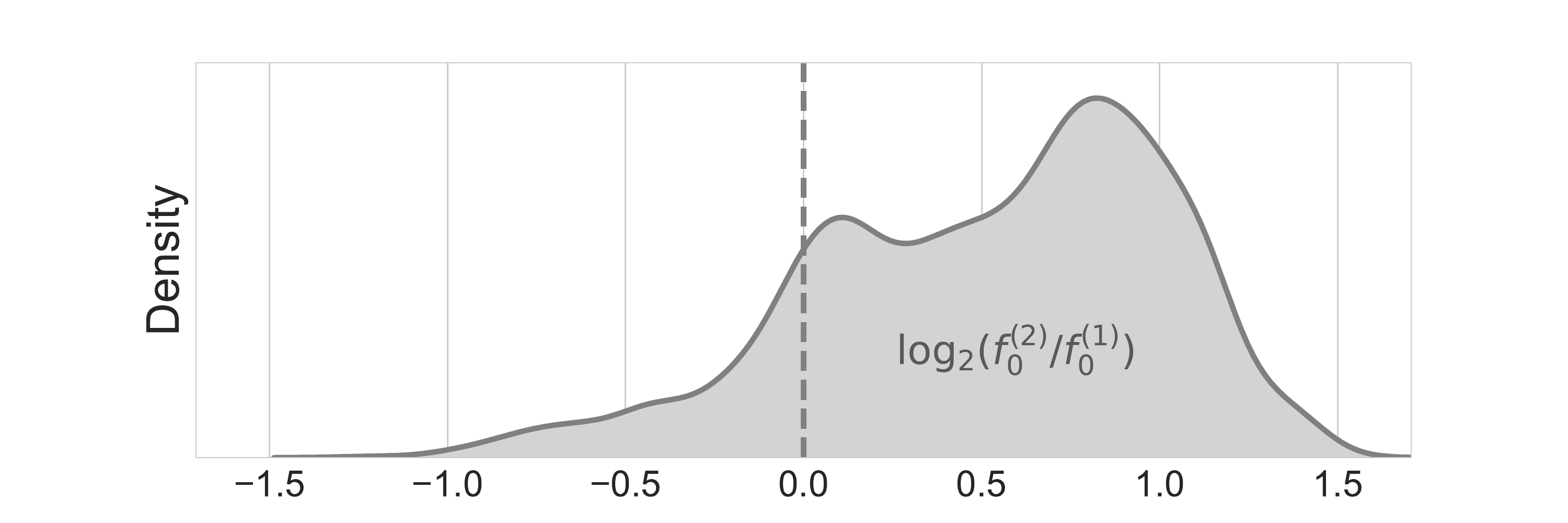}
  \caption{Density over evaluation mixtures of the ratio $\log_2(f_0^{(2)} / f_0^{(1)})$ where $f_0^{(1)}$ and $f_0^{(2)}$ are the average F0s of ConvTasnet's output channels 1 and 2 respectively.}
  \label{fig:f0_bias}
\end{figure}
\subsubsection{Sensitivity of Networks to Temporal Discontinuities}
\vspace*{-2mm}
Our experiments in Sec \ref{sec:continuity_in_time} demonstrate that E2E models quickly lose track of the speaker identities after short silences.
This may be due to the difficulty of the assignment problem: even after performing speaker grouping, the model needs to decide at every time frame which speaker it should assign to a given output channel.
This has been addressed in the past as the label permutation problem - networks are now trained using utterance-level permutation invariant training (U-PIT) \cite{kolbaek2017multitalker, yu2017permutation}, which makes the loss invariant to the order of speakers in the output. However, although the chosen order does not affect the loss function, the network still needs to assign each speaker to a channel in a way that is stable over time. In feed-forward networks, this is decided without any information about past assignments. An option is to compute a statistic that can reliably order the speakers in a temporally consistent way.
We show in Fig. \ref{fig:f0_bias} the distribution of $\log_2(f_0^{(2)} / f_0^{(1)})$ where $f_0^{(i)}$ is the average F0 estimated from the output channel $i$. This distribution is computed using the WSJ-2-mix test-set. It is strongly biased towards positive values, with one peak just above 0 and one near 1, corresponding to mixtures of speakers with the same or different genders respectively. We observe that for $86\%$ of the mixtures, the lower frequency speaker is assigned to channel $1$. 
The network seems to solve the assignment issue by ordering the speakers' fundamental frequencies. Interestingly, this learned trait is in alignment to pre-defined conventions used in early segregation models \cite{weng2015deep}, without it being explicitly constrained during training.
When unsure, the networks seem to assign almost all the energy to the channel 1, explaining the behavior at points of temporal discontinuity. Thus, they rely on the self-consistency of the speakers to perform segregation.
\vspace*{-2mm}
\section{Conclusion}
\vspace*{-1mm}
Our work investigates the performance of E2E speech segregation networks to data with harmonic deformations and temporal discontinuities. Given their poor performance for speech missing the lower and intermediate harmonics, we believe that these models strongly rely on cues from the lower harmonics to detect the harmonic structure.

The network's sensitivity to silences in the mixtures indicates that it enforces consistency of its output over a time window much shorter than its perceptive field. The network assigns speakers to different channels by comparing short-term features (\textit{e.g.} pitch) and is only consistent over time because of the data's consistency.

We also demonstrate that spectrogram based models are more robust to the above deformations, indicating that ConvTasnet's vulnerabilities are not a result of a dataset that prevents \textit{out-of-distribution} generalization but rather a consequence of the learned time-frequency representation. Given that the performance of E2E models have surpassed human performance in ideal conditions, their next challenge is to generalize to speech with such deformations. We believe that using transfer learning \cite{weiss2016survey} or expert-knowledge in designing representations that are more invariant to common deformations, and data-augmentation during training may help bridge this gap in robustness.
\vspace*{-2mm}
\section{Acknowledgements}
\vspace*{-2mm}
This work was supported by NSF grant \#1764010 and an AFOSR grant. The authors declare no conflict of interests.
\bibliographystyle{IEEEtran}

\bibliography{template}

\begin{thebibliography}{10}
\providecommand{\url}[1]{#1}
\csname url@samestyle\endcsname
\providecommand{\newblock}{\relax}
\providecommand{\bibinfo}[2]{#2}
\providecommand{\BIBentrySTDinterwordspacing}{\spaceskip=0pt\relax}
\providecommand{\BIBentryALTinterwordstretchfactor}{4}
\providecommand{\BIBentryALTinterwordspacing}{\spaceskip=\fontdimen2\font plus
\BIBentryALTinterwordstretchfactor\fontdimen3\font minus
  \fontdimen4\font\relax}
\providecommand{\BIBforeignlanguage}[2]{{%
\expandafter\ifx\csname l@#1\endcsname\relax
\typeout{** WARNING: IEEEtran.bst: No hyphenation pattern has been}%
\typeout{** loaded for the language `#1'. Using the pattern for}%
\typeout{** the default language instead.}%
\else
\language=\csname l@#1\endcsname
\fi
#2}}
\providecommand{\BIBdecl}{\relax}
\BIBdecl

\bibitem{parikh2022harmonicity}
R.~Parikh, I.~Kavalerov, C.~Espy-Wilson, and S.~Shamma, ``Harmonicity plays a
  critical role in dnn based versus in biologically-inspired monaural speech
  segregation systems,'' \emph{arXiv preprint arXiv:2203.04420}, 2022.

\bibitem{luo2019conv}
Y.~Luo and N.~Mesgarani, ``Conv-tasnet: Surpassing ideal time--frequency
  magnitude masking for speech separation,'' \emph{IEEE/ACM transactions on
  audio, speech, and language processing}, vol.~27, no.~8, pp. 1256--1266,
  2019.

\bibitem{chen2020dual}
J.~Chen, Q.~Mao, and D.~Liu, ``Dual-path transformer network: Direct
  context-aware modeling for end-to-end monaural speech separation,''
  \emph{arXiv preprint arXiv:2007.13975}, 2020.

\bibitem{hershey2016deep}
J.~R. Hershey, Z.~Chen, J.~Le~Roux, and S.~Watanabe, ``Deep clustering:
  Discriminative embeddings for segmentation and separation,'' in \emph{2016
  IEEE International Conference on Acoustics, Speech and Signal Processing
  (ICASSP)}.\hskip 1em plus 0.5em minus 0.4em\relax IEEE, 2016, pp. 31--35.

\bibitem{zhang2016deep}
X.-L. Zhang and D.~Wang, ``A deep ensemble learning method for monaural speech
  separation,'' \emph{IEEE/ACM transactions on audio, speech, and language
  processing}, vol.~24, no.~5, pp. 967--977, 2016.

\bibitem{luo2018tasnet}
Y.~Luo and N.~Mesgarani, ``Tasnet: time-domain audio separation network for
  real-time, single-channel speech separation,'' in \emph{2018 IEEE
  International Conference on Acoustics, Speech and Signal Processing
  (ICASSP)}.\hskip 1em plus 0.5em minus 0.4em\relax IEEE, 2018, pp. 696--700.

\bibitem{luo2020dual}
Y.~Luo, Z.~Chen, and T.~Yoshioka, ``Dual-path rnn: efficient long sequence
  modeling for time-domain single-channel speech separation,'' in \emph{ICASSP
  2020-2020 IEEE International Conference on Acoustics, Speech and Signal
  Processing (ICASSP)}.\hskip 1em plus 0.5em minus 0.4em\relax IEEE, 2020, pp.
  46--50.

\bibitem{kolbaek2017multitalker}
M.~Kolb{\ae}k, D.~Yu, Z.-H. Tan, and J.~Jensen, ``Multitalker speech separation
  with utterance-level permutation invariant training of deep recurrent neural
  networks,'' \emph{IEEE/ACM Transactions on Audio, Speech, and Language
  Processing}, vol.~25, no.~10, pp. 1901--1913, 2017.

\bibitem{huang2015joint}
P.-S. Huang, M.~Kim, M.~Hasegawa-Johnson, and P.~Smaragdis, ``Joint
  optimization of masks and deep recurrent neural networks for monaural source
  separation,'' \emph{IEEE/ACM Transactions on Audio, Speech, and Language
  Processing}, vol.~23, no.~12, pp. 2136--2147, 2015.

\bibitem{isik2016single}
Y.~Isik, J.~L. Roux, Z.~Chen, S.~Watanabe, and J.~R. Hershey, ``Single-channel
  multi-speaker separation using deep clustering,'' \emph{arXiv preprint
  arXiv:1607.02173}, 2016.

\bibitem{chen2017deep}
Z.~Chen, Y.~Luo, and N.~Mesgarani, ``Deep attractor network for
  single-microphone speaker separation,'' in \emph{2017 IEEE International
  Conference on Acoustics, Speech and Signal Processing (ICASSP)}.\hskip 1em
  plus 0.5em minus 0.4em\relax IEEE, 2017, pp. 246--250.

\bibitem{erdogan2015phase}
H.~Erdogan, J.~R. Hershey, S.~Watanabe, and J.~Le~Roux, ``Phase-sensitive and
  recognition-boosted speech separation using deep recurrent neural networks,''
  in \emph{2015 IEEE International Conference on Acoustics, Speech and Signal
  Processing (ICASSP)}.\hskip 1em plus 0.5em minus 0.4em\relax IEEE, 2015, pp.
  708--712.

\bibitem{shi2019deep}
Z.~Shi, H.~Lin, L.~Liu, R.~Liu, J.~Han, and A.~Shi, ``Deep attention gated
  dilated temporal convolutional networks with intra-parallel convolutional
  modules for end-to-end monaural speech separation.'' in \emph{Interspeech},
  2019, pp. 3183--3187.

\bibitem{narayanan2013ideal}
A.~Narayanan and D.~Wang, ``Ideal ratio mask estimation using deep neural
  networks for robust speech recognition,'' in \emph{2013 IEEE International
  Conference on Acoustics, Speech and Signal Processing}.\hskip 1em plus 0.5em
  minus 0.4em\relax IEEE, 2013, pp. 7092--7096.

\bibitem{kavalerov2019universal}
I.~Kavalerov, S.~Wisdom, H.~Erdogan, B.~Patton, K.~Wilson, J.~Le~Roux, and
  J.~R. Hershey, ``Universal sound separation,'' in \emph{2019 IEEE Workshop on
  Applications of Signal Processing to Audio and Acoustics (WASPAA)}.\hskip 1em
  plus 0.5em minus 0.4em\relax IEEE, 2019, pp. 175--179.

\bibitem{defossez2019music}
A.~D{\'e}fossez, N.~Usunier, L.~Bottou, and F.~Bach, ``Music source separation
  in the waveform domain,'' \emph{arXiv preprint arXiv:1911.13254}, 2019.

\bibitem{Wisdom_InPrep2020}
S.~Wisdom, H.~Erdogan, D.~P.~W. Ellis, R.~Serizel, N.~Turpault, E.~Fonseca,
  J.~Salamon, P.~Seetharaman, and J.~R. Hershey, ``What's all the fuss about
  free universal sound separation data?'' in \emph{in preparation}, 2020.

\bibitem{Turpault2019_DCASE}
\BIBentryALTinterwordspacing
N.~Turpault, R.~Serizel, A.~Parag~Shah, and J.~Salamon, ``{Sound event
  detection in domestic environments with weakly labeled data and soundscape
  synthesis},'' in \emph{{Workshop on Detection and Classification of Acoustic
  Scenes and Events}}, New York City, United States, October 2019. [Online].
  Available: \url{https://hal.inria.fr/hal-02160855}
\BIBentrySTDinterwordspacing

\bibitem{le2019sdr}
J.~Le~Roux, S.~Wisdom, H.~Erdogan, and J.~R. Hershey, ``Sdr--half-baked or well
  done?'' in \emph{ICASSP 2019-2019 IEEE International Conference on Acoustics,
  Speech and Signal Processing (ICASSP)}.\hskip 1em plus 0.5em minus
  0.4em\relax IEEE, 2019, pp. 626--630.

\bibitem{mcfee2015librosa}
B.~McFee, C.~Raffel, D.~Liang, D.~P. Ellis, M.~McVicar, E.~Battenberg, and
  O.~Nieto, ``librosa: Audio and music signal analysis in python,'' in
  \emph{Proceedings of the 14th python in science conference}, vol.~8.\hskip
  1em plus 0.5em minus 0.4em\relax Citeseer, 2015, pp. 18--25.

\bibitem{vincent2006performance}
E.~Vincent, R.~Gribonval, and C.~F{\'e}votte, ``Performance measurement in
  blind audio source separation,'' \emph{IEEE transactions on audio, speech,
  and language processing}, vol.~14, no.~4, pp. 1462--1469, 2006.

\bibitem{weiss2010speech}
R.~J. Weiss and D.~P. Ellis, ``Speech separation using speaker-adapted
  eigenvoice speech models,'' \emph{Computer Speech \& Language}, vol.~24,
  no.~1, pp. 16--29, 2010.

\bibitem{cooke2010monaural}
M.~Cooke, J.~R. Hershey, and S.~J. Rennie, ``Monaural speech separation and
  recognition challenge,'' \emph{Computer Speech \& Language}, vol.~24, no.~1,
  pp. 1--15, 2010.

\bibitem{elhilali2008cocktail}
M.~Elhilali and S.~A. Shamma, ``A cocktail party with a cortical twist: how
  cortical mechanisms contribute to sound segregation,'' \emph{The Journal of
  the Acoustical Society of America}, vol. 124, no.~6, pp. 3751--3771, 2008.

\bibitem{krishnan2014segregating}
L.~Krishnan, M.~Elhilali, and S.~Shamma, ``Segregating complex sound sources
  through temporal coherence,'' \emph{PLoS computational biology}, vol.~10,
  no.~12, p. e1003985, 2014.

\bibitem{vishnubhotla2009algorithm}
S.~Vishnubhotla and C.~Y. Espy-Wilson, ``An algorithm for speech segregation of
  co-channel speech,'' in \emph{2009 IEEE International Conference on
  Acoustics, Speech and Signal Processing}.\hskip 1em plus 0.5em minus
  0.4em\relax IEEE, 2009, pp. 109--112.

\bibitem{stark2010source}
M.~Stark, M.~Wohlmayr, and F.~Pernkopf, ``Source--filter-based single-channel
  speech separation using pitch information,'' \emph{IEEE Transactions on
  Audio, Speech, and Language Processing}, vol.~19, no.~2, pp. 242--255, 2010.

\bibitem{wang1999separation}
D.~L. Wang and G.~J. Brown, ``Separation of speech from interfering sounds
  based on oscillatory correlation,'' \emph{IEEE transactions on neural
  networks}, vol.~10, no.~3, pp. 684--697, 1999.

\bibitem{chi2005multiresolution}
T.~Chi, P.~Ru, and S.~A. Shamma, ``Multiresolution spectrotemporal analysis of
  complex sounds,'' \emph{The Journal of the Acoustical Society of America},
  vol. 118, no.~2, pp. 887--906, 2005.

\bibitem{shamma2000case}
S.~Shamma and D.~Klein, ``The case of the missing pitch templates: how harmonic
  templates emerge in the early auditory system,'' \emph{The Journal of the
  Acoustical Society of America}, vol. 107, no.~5, pp. 2631--2644, 2000.

\bibitem{yu2017permutation}
D.~Yu, M.~Kolb{\ae}k, Z.-H. Tan, and J.~Jensen, ``Permutation invariant
  training of deep models for speaker-independent multi-talker speech
  separation,'' in \emph{2017 IEEE International Conference on Acoustics,
  Speech and Signal Processing (ICASSP)}.\hskip 1em plus 0.5em minus
  0.4em\relax IEEE, 2017, pp. 241--245.

\bibitem{weng2015deep}
C.~Weng, D.~Yu, M.~L. Seltzer, and J.~Droppo, ``Deep neural networks for
  single-channel multi-talker speech recognition,'' \emph{IEEE/ACM Transactions
  on Audio, Speech, and Language Processing}, vol.~23, no.~10, pp. 1670--1679,
  2015.

\bibitem{weiss2016survey}
K.~Weiss, T.~M. Khoshgoftaar, and D.~Wang, ``A survey of transfer learning,''
  \emph{Journal of Big data}, vol.~3, no.~1, pp. 1--40, 2016.

\end{thebibliography}


\end{document}